\documentclass[letterpaper, 10 pt, conference]{ieeeconf}
\IEEEoverridecommandlockouts
\let\labelindent\relax

\usepackage{graphicx}
\usepackage{tikz}
\usetikzlibrary{backgrounds}
\usetikzlibrary{positioning,fit,calc,decorations.pathreplacing,arrows.meta}
\usepackage{pgfplots}
\usepackage{pgfplotstable}
\usepackage{stfloats} 
\usetikzlibrary{pgfplots.groupplots,pgfplots.statistics,pgfplots.fillbetween}
\pgfplotsset{compat=1.18}
\usepackage{adjustbox}
\usepackage{cite}

\usepackage{amsmath,amssymb,mathtools,amsthm,aligned-overset}
\usepackage{algorithmic}
\usepackage{hyperref}
\usepackage{xcolor}
\usepackage{enumitem}
\newtheorem{theorem}{Theorem}

\newtheorem{assumption}{Assumption}

\newtheorem*{remark}{Remark}

\DeclareMathOperator{\Var}{Var}

\usepackage{booktabs}
\usepackage{array}
\usepackage[dvipsnames,table,xcdraw]{xcolor}

\usepackage[acronym]{glossaries}
\glsdisablehyper
\newacronym{ccd}{CCD}{Charge Coupled Device}
\newacronym{piv}{PIV}{Particle Image Velocimetry}
\newacronym{epe}{EPE}{end-point error}
\newacronym{nll}{NLL}{negative log-likelihood}
\newacronym{dis}{DIS}{Dense Inverse Search}
\newacronym{log}{LoG}{Laplacian-of-Gaussian}
\newacronym{blue}{BLUE}{Best Linear Unbiased Estimator}
\newacronym{of}{OF}{Optical Flow}
\newacronym{map}{MAP}{Maximum A Posteriori}
\newacronym{admm}{ADMM}{Alternating Direction Method of Multipliers}

\newcommand{\startflow}{\hat{\mathbf{u}}}
\newcommand{\optConsensusVar}{\mathbf{u}}
\newcommand{\optVar}{\mathbf{u}}
\newcommand{\numAgents}{N}

\newcommand{\groundtruthflow}{{\mathbf{u}}^\ast}
\newcommand{\dual}{\mathbf{d}}

\newcommand{\numPixels}{n}

\definecolor{leftbg}{RGB}{237,246,255}   
\definecolor{midbg}{RGB}{255,243,230}    
\definecolor{rightbg}{RGB}{235,246,238}  
\definecolor{listbg}{RGB}{236,241,255}   
\definecolor{textgray}{RGB}{60,60,60}

\newcounter{myalgorithm}
\newenvironment{myalgorithm}[1][]
{	\refstepcounter{myalgorithm}
	\begin{minipage}{\linewidth}
		\medskip
		\hrule
		\smallskip
		\textsc{Algorithm \themyalgorithm}. #1
		\smallskip
		\hrule 
		\smallskip
	\end{minipage}
}
{
	\smallskip
	\hrule width\linewidth\relax
	\smallskip
}

\DeclareMathOperator{\argmin}{argmin}
\usepackage{mdframed}

\usepackage[capitalize]{cleveref}
\crefname{example}{Example}{Examples}
\crefname{proposition}{Proposition}{Propositions}
\crefname{myalgorithm}{Algorithm}{Algorithms}
\crefname{assumption}{Assumption}{Assumptions}
\crefname{appendix}{Appendix}{Appendices}
\crefname{assumptionenumi}{}{}
\newtheorem{proposition}[theorem]{Proposition}

\theoremstyle{definition}

\usepackage{tcolorbox}

\title{\LARGE \bf Particle Image Velocimetry Refinement\\via Consensus ADMM for Active Fluid Control}
\author{Alan Bonomi$^\ast$, Francesco Banelli$^\ast$ and Antonio Terpin$^\ast$
\thanks{$^\ast$: All authors contributed equally to this work and are with the Institute for Dynamic Systems and Control, ETH Zurich.}
\thanks{Corresponding author: Antonio Terpin ({\tt\small aterpin@ethz.ch}).}}

\begin{document}

\maketitle
\thispagestyle{empty}
\pagestyle{empty}

\begin{abstract}
Particle Image Velocimetry (PIV) is among the central modalities for measuring flow fields across laboratory, industrial and environmental setting. Traditional PIV approaches typically depend on tuning parameters specific to the imaging setup, making the performance sensitive to variations in illumination, flow conditions, and seeding density. Similarly, state-of-the-art machine learning methods for flow quantification are fragile outside their training set.
In our experiments, we observed that flow quantification would improve if different tunings (or algorithms) were applied to different regions of the same image pair.
Motivated by this observation, we thus pose flow quantification as a multi-estimator fusion problem: several heterogeneous algorithms process the same image pair in parallel, and their dense flow fields are treated as complementary estimates. To fuse them, we adopt a consensus framework based on the alternating direction method of multipliers, incorporating priors such as smoothness and incompressibility.
We perform several numerical experiments to demonstrate the benefits of this approach. For instance, we achieve a decrease in end-point-error of up to 20~\% of a dense-inverse-search estimator at an inference rate of 60~Hz, and we show how performance can be increased with outlier rejection.
Our method is implemented in JAX and integrated into Flow Gym, enabling reproducible comparisons with the state of the art and systematic evaluation across different base algorithms. 
Finally, we demonstrate successful deployment of our method in the same real-world active-fluids-control setup of Terpin and D'Andrea~\cite{terpin2025ff}, where a reinforcement-learning agent uses our flow estimates to learn to minimize drag (down by $36~\%$) or maximize it (up to $32~\%$) with only two minutes of real-world interaction.
Hardware and software are made available at \url{ActiveFluidControl.com}.
\end{abstract}
\section{Introduction}
\label{sec:intro}
\gls*{piv} \cite{Adrian1991,willert2007particle,westerweel2013particle} and related image-based flow-quantification methods \cite{Liu2008,corpetti2006fluid,zhong2017optical} are classical tools for measuring flow fields across experimental fluid mechanics, with applications spanning laboratory \cite{Adrian1991}, industrial \cite{willert2007particle}, microfluidic \cite{Santiago1998,Meinhart1999}, and environmental settings \cite{Muste2008,legleiter2024framework}. Traditionally, their role has been primarily observational: to reconstruct velocity fields for analysis, validation, and diagnostics. For this reason, performance has classically been assessed mainly through accuracy, often using the \gls*{epe}, i.e., the average $\ell_2$ norm of the difference between the ground truth and the estimated value of the flow field.

Recent results in active fluid control motivate a broader perspective. In their recent study, Terpin and D'Andrea \cite{terpin2025ff} showed that, in a real-world fluids-control task, an agent equipped with dense flow feedback could discover high-performance strategies within minutes of real-world interaction, whereas removing that feedback caused learning to fail altogether.
\begin{figure}
    \centering
    \includegraphics[width=\linewidth]{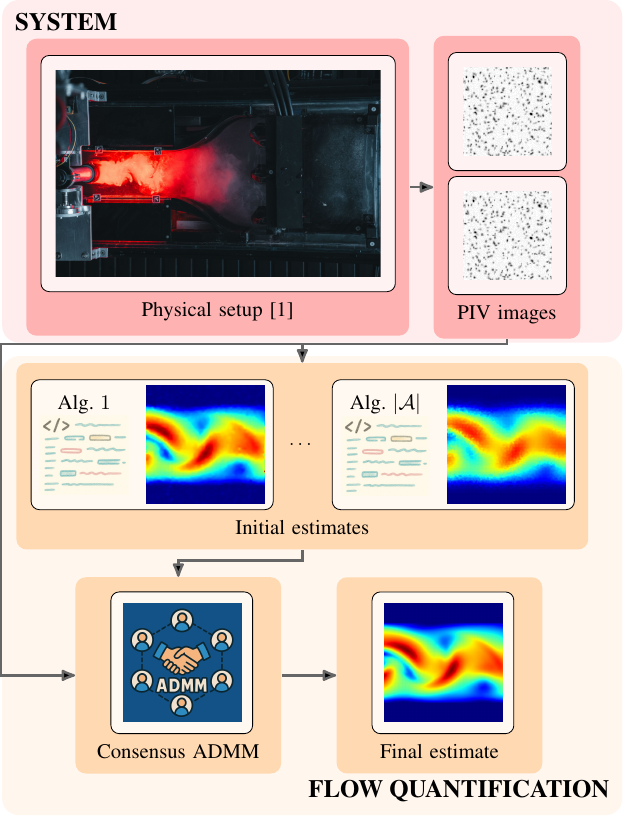}
    \caption{The proposed consensus-ADMM pipeline for PIV refinement. Consecutive PIV images of illuminated tracer particles are processed by multiple estimators in parallel, and their outputs are fused through consensus ADMM into a refined flow field.}
    \label{fig:pipeline}
    \vspace{-.5cm}
\end{figure}
This observation introduces a different axis in the evaluation of PIV algorithms. Once flow estimation is embedded in a real-time learning or control loop, latency becomes equally central. A highly accurate but slow estimator cannot run online, whereas a fast estimator with poor accuracy may provide feedback that is too crude to support learning. In this regime, the relevant objective is the speed--accuracy trade-off. 
This tension is already visible in current methods: using Flow Gym \cite{BANELLI2026102641}, one can verify that the state-of-the-art RAFT32-PIV \cite{cai2019dense} reaches only up to 1.7\,Hz on an RTX 4090 GPU, while fast estimators such as DIS \cite{kroeger2016fast} can operate at 60\,Hz or more, but at substantially lower accuracy.

\begin{figure}
    \centering
    \begin{minipage}{.32\linewidth}
    \centering
        \includegraphics[width=\linewidth]{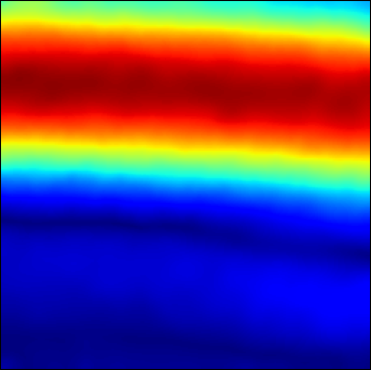}
    \end{minipage}
\begin{minipage}{.32\linewidth}
        \centering
        \includegraphics[width=\linewidth]{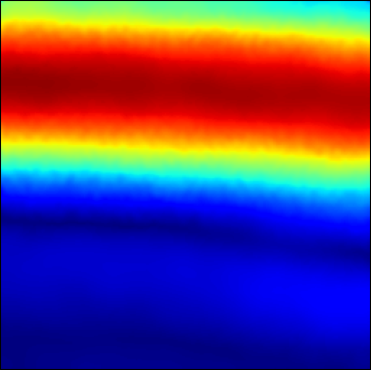}
    \end{minipage}
    \hfill
    \begin{minipage}{.32\linewidth}
    \centering
        \includegraphics[width=\linewidth]{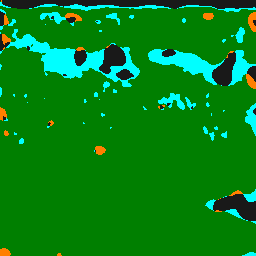}
    \end{minipage}
    \caption{On the left and middle, we report the flow estimates (colors represent the magnitude of the flow field at each pixel) for the same \gls*{piv} image pair of two different tunings of the JAX DIS implementation from Flow Gym \cite{BANELLI2026102641}. On the right, we mark in green pixels in which both estimates have an estimate closer than $0.1\mathrm{px}$ to the ground-truth, light blue if only the first tuning, orange if only the second, and black if both estimates fail to be closer than $0.1\mathrm{px}$ to the ground-truth.}
    \label{fig:dep-parameters}
    \vspace{-.5cm}
\end{figure}

There are, broadly speaking, two ways to improve this operating point: one may attempt to accelerate highly accurate but computationally expensive estimators, or instead improve the accuracy of methods that already satisfy the real-time constraint. In this work, we pursue the latter direction. A central obstacle is that no single estimator or parameter setting performs uniformly best across all flows, or even across all regions of the same image pair. In practice, classical flow estimators often require parameter choices that depend on the experimental setup \cite{kahler_main_2016}, and recent work has begun to explore dynamic parameter adaptation to mitigate this sensitivity \cite{jassal2025optical}. From our experience, it is often easier to tune an estimator for a specific regime than to identify a single configuration that performs well across heterogeneous conditions.
This effect can arise even within a single flow field. In \cref{fig:dep-parameters}, we compare two parameter tunings of DIS on the same PIV image pair and observe that each tuning is more accurate on different spatial regions. This suggests that robustness may be improved not by searching for a single universal estimator, but by combining complementary candidate estimates and enforcing global consistency across the resulting field.

\begin{mdframed}[hidealllines=true,backgroundcolor=blue!5]
\noindent\textbf{Contributions.}
Motivated by this observation, we pose flow quantification as a multi-estimator fusion problem: several heterogeneous
algorithms process the same image pair, and their dense velocity fields
are treated as complementary observations. To the best of our knowledge, this is the first formulation and experimental demonstration of parallel
estimator-output fusion for fluid velocimetry.
We combine the different estimates via consensus \gls*{admm}, enforcing priors such as smoothness and incompressibility; \cref{fig:pipeline}. In this paper, we showcase the results of this approach using as base algorithms DIS \cite{kroeger2016fast}, DeepFlow \cite{weinzaepfel2013deepflow}, and Farneb\"ack \cite{farneback2003two}, but the method applies to any other algorithm, possibly learning-based. For instance, with DIS, we achieve a decrease in \gls*{epe} of up to $20~\%$ at an inference rate of $60~\mathrm{Hz}$, and we show how performance can be increased with outlier-rejection. 
Our method is implemented in JAX and integrated into Flow Gym \cite{BANELLI2026102641}, enabling reproducible evaluation and direct use in downstream control tasks. In particular, we deploy it on the real-world active-fluids-control setup of \cite{terpin2025ff}, where a reinforcement-learning agent uses our flow estimates to learn to minimize drag (down by $36~\%$) or maximize it (up to $32~\%$) with only two minutes of real-world interaction.
\end{mdframed}

\paragraph*{Notation}
We let $\numPixels$ be the number of pixels. The single-channel input frames are $I_0, I_1$. Since $\Delta t$ and the pixel-to-world conversion are known, estimating velocity is equivalent to estimating the pixel displacement $(u_x,u_y)\in\mathbb{R}^2$ at each pixel. We denote the full flow field by $\optVar=(\optVar_x,\optVar_y)\in\mathbb{R}^{2\numPixels}$, with entries indexed by $(\cdot)_\ell$, $\ell\in\{1,\dots,2\numPixels\}$.
\section{PIV refinement}
\label{sec:problem}
In \gls*{piv}, flow estimation is commonly formulated as the minimization of an image-consistency term together with a regularizer. Under the usual small-displacement assumption, the short inter-frame delay often makes brightness constancy a reasonable approximation, i.e., $I_0(x,y)\approx I_1(x+u_x,y+u_y)$. This leads to objectives of the form
\begin{equation}
    \underset{\optConsensusVar \in \mathbb{R}^{2\numPixels}}{\mathrm{min}} \;
    \mathcal{J}_d(\optConsensusVar,(I_0,I_1))
    + \lambda \mathcal{J}_r(\optConsensusVar,(I_0,I_1)),
    \label{eq:optical_flow_objective}
\end{equation}
where $\mathcal{J}_d$ measures consistency with the image pair and $\mathcal{J}_r$ encodes priors such as smoothness.

In our setting, however, we are not interested in solving \eqref{eq:optical_flow_objective} from scratch with a single estimator. Instead, given a collection of \gls*{piv} algorithms $\mathcal{A}=\{A_1,\ldots,A_\numAgents\}$ producing candidate flows $\startflow_i=A_i(I_0,I_1)\in\mathbb{R}^{2\numPixels}$, we seek a refined field that remains close to these candidates where they are reliable, while enforcing global priors. This leads to the optimization problem
\begin{align}
    \underset{\optConsensusVar \in \mathbb{R}^{2\numPixels}}{\mathrm{min}} \;
    \sum_{i=1}^{\numAgents}
    \underbrace{\mathcal{J}_i(\optConsensusVar,A_i(I_0,I_1),(I_0,I_1))}_{=:f_i(\optConsensusVar,(I_0,I_1))}
    + g(\optConsensusVar,(I_0,I_1)),
    \label{eq:optimization_framework}
\end{align}
where each $f_i$ measures agreement with the $i$-th candidate estimate, and $g$ collects global regularization terms expressing priors such as smoothness and incompressibility.

\subsection{PIV refinement via Consensus ADMM}
\label{sec:method}

In \eqref{eq:optimization_framework}, each data term $f_i(\optConsensusVar, (I_0, I_1))$ is defined by the individual algorithm estimate $\startflow_i = A_i(I_0, I_1)$ and by the image pair $(I_0, I_1)$. By equivalently writing \eqref{eq:optimization_framework} as
\begin{subequations}\label{eq:distributed_optimization_framework}
\begin{align}
    \underset{\optVar_1,...,\optVar_{\numAgents}, \optConsensusVar \in \mathbb{R}^{2\numPixels}}{\mathrm{min}} \quad & 
    \sum_{i=1}^\numAgents f_i(\optVar_i, (I_0, I_1)) + g(\optConsensusVar, (I_0, I_1)) \label{eq:dist_opt_a} \\
    \text{s.t.} \quad & 
    \optVar_i - \optConsensusVar = 0, \quad i = 1, \dots, \numAgents \label{eq:dist_opt_b}
\end{align}
\end{subequations}
we match the formulation in \cite[Problem~7.2]{stephen_boyd_distributed_2011}. The coupling constraints 
$\optVar_i = \optConsensusVar, i \in \{1, \dots, N\}$
enforce that all local estimates $\optVar_i$ agree on a single consensus field $\optConsensusVar$. 
With the dual variables $\dual_i \in \mathbb{R}^{2\numPixels}$, the augmented Lagrangian reads
$
\mathcal{L}_\rho(\optVar_1, \ldots, \optVar_{\numAgents}, \dual_1, \ldots, \dual_N, \optConsensusVar) 
=
\sum_{i=1}^N f_i(\optVar_i, (I_0, I_1)) + g(\optConsensusVar, (I_0, I_1))
+ \dfrac{\rho}{2} \sum_{i=1}^N \|\optVar_i - \optConsensusVar + \dual_i\|_2^2,
$
where $\rho > 0$ is the penalty parameter; in our experiments, we use $\rho = 2$.
\begin{remark}[Penalty-parameter sensitivity]
\gls{admm} performance can be sensitive to the penalty parameter $\rho$. Under standard convergence assumptions \cite[Section~3.2]{stephen_boyd_distributed_2011}, $\rho$ does not change the limiting solution, but it can affect the convergence rate: small values may weakly enforce consensus, whereas large values may overemphasize the proximal term relative to data fidelity. Since the relative scales of $f_i$ and $g$ may vary across heterogeneous flow regimes, the adaptive selection of $\rho$ is a promising direction for future work.
\end{remark}
By applying the \emph{global consensus} variant of \gls{admm} \cite[Section~7.2]{stephen_boyd_distributed_2011} we obtain the fixed-point iteration\footnote{We use the square brackets to denote the value of the same quantity at different iterations.}:
\begin{subequations} \label{eq:dra_for_projection}
\begin{align}
&\begin{aligned}
\optVar_i[k + 1] &= \underset{\optVar_i \in \mathbb{R}^{2\numPixels}}{\argmin}\; f_i(\optVar_i, (I_0, I_1))
\\&+ \dfrac{\rho}{2} \|\optVar_i - \optConsensusVar[k] + \dual_i[k]\|_2^2,
\end{aligned}
\label{eq:local-update}
\\
&\begin{aligned}
    \optConsensusVar[k+1] &= \underset{\optConsensusVar \in \mathbb{R}^{2\numPixels}}{\argmin}\; g(\optConsensusVar, (I_0, I_1)) 
\\&+ \dfrac{\rho}{2} \sum_{i=1}^N \|\optConsensusVar - \optVar_i[k+1] -\dual_i[k]\|_2^2,
\end{aligned}
\label{eq:global-update} 
\\
&\begin{aligned}
\dual_i[k+1] &= \dual_i[k] + \optVar_i[k + 1] - \optConsensusVar[k + 1],
\end{aligned}
\label{eq:dual-update}
\end{align}
\end{subequations}
where the updates \eqref{eq:local-update} and \eqref{eq:dual-update} are for all $i \in \{1, \ldots, \numAgents\}$.
In the remainder of this section, we show that for $f_i(\cdot)$ and $g(\cdot)$ of interest for the \gls*{piv} refinement problem in \eqref{eq:optimization_framework}, we can efficiently compute the iterations \eqref{eq:local-update}-\eqref{eq:global-update}, enabling the implementation of \cref{algo:consensus_admm}:

\noindent\begin{myalgorithm}[PIV Refinement via Consensus ADMM] \label{algo:consensus_admm}%
    \textbf{Inputs:} $\startflow_i \in \mathbb{R}^{2\numPixels}$ $\forall i \in \{1, \ldots, \numAgents\}$.\\
    \textbf{Initialization:} $\optConsensusVar[0] = \dfrac{1}{\numAgents}\sum_{i=1}^{\numAgents}\startflow_i$\\
    $\optVar_i[0]= \startflow_i$ $\forall i \in \{1, \ldots, \numAgents\}$\\
    $\dual_i[0] = \startflow_i - \optConsensusVar[0]$ $\forall i \in \{1, \ldots, \numAgents\}$\\
    \textbf{For $k = 0$ to $K_1 - 1$:} \\
    $
    \left\lfloor
    \begin{array}{l}
           \text{Update }\optVar_i[k+1]\text{ according to \eqref{eq:local-via-proximal} in \cref{subsec:distributed_data_term}.}\\
           \bar \optVar[k+1] \gets \dfrac{1}{\numAgents}\sum_{i=1}^\numAgents \optVar_i[k+1]\\
           \bar \dual[k] \gets \dfrac{1}{\numAgents}\sum_{i=1}^\numAgents \dual_i[k]\\
           \optConsensusVar[k+1] \gets \text{\cref{algo:global_update}}(\bar \optVar[k+1], \bar \dual[k], \optConsensusVar[k])\\
           \dual_i[k+1] \gets \dual_i[k] + \optVar_i[k+1] - \optConsensusVar[k+1]\\[.3em]
        
    \end{array}
    \right.
    $ \\[.3em]
    \textbf{Output:} $\optConsensusVar[k+1]$
\end{myalgorithm}%
Standard \gls*{admm} convergence results apply to the exact convex updates in \cref{algo:consensus_admm} (see, e.g., \cite[Section~3.2]{stephen_boyd_distributed_2011}). In the real-time implementation considered here, we instead use a fixed iteration budget of $K_1 = 30$. 
A systematic analysis of the sensitivity of performance to $K_1$ is omitted for brevity, and we believe that future work may address the applicability of the method at higher frame rates or on lower-power hardware, as well as how performance can improve with more computational resources.

\subsection{Data term and \texorpdfstring{update \eqref{eq:local-update}}{local update}.}
\label{subsec:distributed_data_term}
An effective choice for $f_i(\cdot)$ is given, for a closed, proper, and convex scalar $\phi(\cdot)$ by 
\begin{equation}
    f_i(\optVar_i, (I_0, I_1)) 
    = \sum_{\ell = 1}^{2\numPixels}\phi(w_{i,\ell}(I_0,I_1)(\optVar_i - \startflow_i)_\ell),
    \label{eq:distributed_data_term}
\end{equation}
where $w_{i,\ell}(I_0,I_1) \geq 0$ encodes the estimate and pixel-wise confidence in $\startflow_i$ and may depend on the image pair $(I_0, I_1)$. 
Hereafter, we will drop the dependency on $I_0, I_1$ for ease of notation in all the quantities depending on $I_0, I_1$. 
Then, the update \eqref{eq:local-update} boils down to solving
\begin{multline*}
\underset{\optVar_i \in \mathbb{R}^{2\numPixels}}{\argmin}
\sum_{\ell = 1}^{2\numPixels}\phi(w_{i,\ell}(\optVar_i - \startflow_i)_\ell)
+ \dfrac{\rho}{2}
\sum_{\ell = 1}^{2\numPixels}(\optVar_i - \optVar[k] + \dual_i[k]\bigr)_\ell^2.
\end{multline*}
Thus, with $c = (\optVar[k] - \dual_i[k])_\ell$ and $r = (\optVar[k] - \startflow_i - \dual_i[k])_\ell$, the update reduces to $2\numPixels\numAgents$ scalar problems:
\begin{align}
&\underset{t \in \mathbb{R}}{\argmin}\;
\phi(w_{i,\ell}(t - (\startflow_i)_\ell))
+ \dfrac{\rho}{2}\bigl(t - c\bigr)^2\notag
\\&=
(\startflow_i)_\ell + \underset{t \in \mathbb{R}}{\argmin}\;
\left(\dfrac{1}{\rho}\phi(w_{i,\ell}t)
+ \dfrac{1}{2}\bigl(t - r\bigr)^2\right).
\label{eq:local-via-proximal}
\end{align}
That is, the update \eqref{eq:local-update} amounts to a per-pixel proximal evaluation. For many $\phi$ of interest, this can be done in closed form. For instance, this is the case for the $\ell_1$ norm, the squared $\ell_2$ norm, and the Huber loss; cf. \cref{appendix:other-losses}. We compare these alternatives in \cref{sec:numerical}.

The weights $w_{i,\ell}$ can be computed in several different ways. In \cref{sec:numerical}, we show that $w_{i,\ell} = 1$ performs well, but we also consider different weighting schemes; cf. also \cref{appendix:confidence}. 
Finally, if the estimates are subject to an outlier-rejection scheme before being provided to our method, one can simply set the weights corresponding to outliers to zero and \eqref{eq:local-update}-\eqref{eq:dual-update} will implicitly perform data interpolation according to the other estimates and priors.

\subsection{Regularization term and \texorpdfstring{update \eqref{eq:global-update}}{global update}.}
\label{subsec:regularization}
We embed some of the priors on the solution via $g(\cdot)$:
\begin{equation}
  g(\optConsensusVar) = \lambda_{\mathrm{s}} \mathcal{R}_{\mathrm{s}}(\optConsensusVar) 
           + \lambda_{\mathrm{curv}}  \mathcal{R}_{\mathrm{curv}}(\optConsensusVar)
           + \lambda_{\mathrm{div}}  \mathcal{R}_{\mathrm{div}}(\optConsensusVar),
  \label{eq:reg_sum}
\end{equation}
where each $\lambda_\bullet \ge 0$ is a tunable parameter. Each penalty encodes a prior about the spatial regularity or structure of the flow field:
\begin{itemize}[leftmargin=*]
    \item \emph{Smoothness}: We penalize sharp variations across estimates of nearby pixels,
    \begin{equation}
    \begin{aligned}
  \mathcal{R}_{\mathrm{s}}(\optConsensusVar) 
  = \|D_x \optConsensusVar_x\|_2^2 + \|D_y \optConsensusVar_x\|_2^2 + \|D_x \optConsensusVar_y\|_2^2 + \|D_y \optConsensusVar_y\|_2^2,
   \end{aligned}
  \label{eq:smooth}
    \end{equation}
    where $D_x, D_y$ are the first-order central finite-difference operators along the $x$- and $y$-directions, evaluated only on the interior grid; accordingly, derivative-based quantities are computed on the cropped domain obtained by excluding one-pixel boundaries.
This penalty term suppresses high spatial frequencies and is straightforward to discretize, but may oversmooth motion discontinuities.
    \item \emph{Second-order spatial variation}: We penalize second-order spatial variations of the flow by convolving (\( * \)) each velocity component with a standard discrete \(3 \times 3\) Laplacian stencil (equivalently, a narrow-scale Laplacian-of-Gaussian ($\mathrm{LoG}$) approximation \cite{marr1980theory}):
\begin{equation}
  \mathcal{R}_{\mathrm{curv}}(\optConsensusVar)
  = \|\mathrm{LoG} * \optConsensusVar_x \|_2^2 + \|\mathrm{LoG} * \optConsensusVar_y \|_2^2.
  \label{eq:log_acc}
\end{equation}
    \item \emph{Incompressibility}: We penalize non-zero values of the divergence of the flow field:
\begin{equation}
  \mathcal{R}_{\mathrm{div}}(\optConsensusVar) 
  = \|\nabla\cdot \optConsensusVar\|_2^2
  = \|D_x \optConsensusVar_x + D_y \optConsensusVar_y\|_2^2.
  \label{eq:div}
\end{equation}
\end{itemize}

The parameters $\lambda_\bullet$ are easy to tune: they represent how much the optimization should weigh certain regularity priors. 
We describe a simple empirical procedure in \cref{appendix:regularizer-tuning}.

With this choice, the update in \eqref{eq:global-update} is a strictly convex quadratic problem, which we solve approximately by running $K_2 = 30$ iterations of gradient descent with the Adam optimizer \cite{kingma2014adam}, using default hyperparameters and step size $\eta = 0.01$. The overall procedure is summarized in \cref{algo:global_update}. 
This design integrates naturally with our JAX pipeline and yields a simple fixed-budget implementation compatible with the real-time setting targeted in this work. At the same time, the optimality condition of \eqref{eq:global-update} defines a sparse structured linear system induced by the finite-difference and convolution stencils appearing in the regularization terms. Exploiting this structure more directly through matrix-free Krylov \cite{saad_iterative_2003}, multigrid methods \cite{briggs2000multigrid}, conjugate
gradient iterations \cite{hestenes1952methods}, and hard-constrained neural networks \cite{grontas2026pinet} is a promising future research direction. 

\noindent
\begin{myalgorithm}[Global update] \label{algo:global_update}%
    \textbf{Choose:} $\eta$\\
    \textbf{Inputs:} $ \bar \optVar[k+1], \bar \dual[k], \optConsensusVar[k]$\\
    \textbf{Initialization:} $\mathbf{z}[0] = \optConsensusVar[k]$\\
    \textbf{For $j = 0$ to $K_2 - 1$:} \\
    $
    \left\lfloor
    \begin{array}{l}
           \nabla_{\mathbf{z}[j]} \mathcal{L}(\mathbf{z}[j]) = \\\nabla_{\mathbf{z}[j]}( g(\mathbf{z}[j]) + \dfrac{\numAgents\rho}{2}\|\mathbf{z}[j] - \bar \optVar[k+1] - \bar \dual[k]\|_2^2)\\
           \mathbf{z}[j + 1] \gets \text{Adam step}\big(\mathbf{z}[j], \nabla_{\mathbf{z}[j]} \mathcal{L}(\mathbf{z}[j]),\eta \big)
    \end{array}
    \right.
    $ \\[.3em]
    \textbf{Output:} $\mathbf{z}[K_2]$
\end{myalgorithm}%

\section{Numerical Results}
\label{sec:numerical}
In this section, we numerically assess the performance of our method. The code to reproduce all the experiments is available in Flow Gym \cite{BANELLI2026102641}. We consider the widely adopted PIV dataset \cite{cai2019dense}, which comprises $256\times256$ image pairs and ground-truth flow data, utilizing SynthPix as the dataloader \cite{TERPIN2026102642}.
For all numerical experiments, we evaluate the proposed method using three base estimators available in Flow Gym \cite{BANELLI2026102641}: DIS \cite{kroeger2016fast}, DeepFlow \cite{weinzaepfel2013deepflow}, and Farneb\"ack \cite{farneback2003two}. For DIS and Farneb"ack, we construct three parameter configurations by tuning each method on $10~\%$ of the training split of \cite{cai2019dense}, with one configuration tuned on each of the following class groups: (Cylinder, JHTDB), (Backstep, Uniform), and (SQG, DNS). We denote the resulting variants by $\mathrm{DIS}1,\mathrm{DIS}2,\mathrm{DIS}3$ and $\mathrm{F}1,\mathrm{F}2,\mathrm{F}3$, respectively. We further denote by $\mathrm{DIS}^\ast$ and $\mathrm{F}^\ast$ the best-performing DIS and Farneb\"ack configurations when evaluated across all categories. Since the current OpenCV implementation \cite{opencv_library} does not expose tunable parameters for DeepFlow, we use a single DeepFlow instance, denoted by $\mathrm{DF}$.
We apply the proposed framework to the estimator collections
$
\mathcal{A} \in \bigl\{
\{\mathrm{DIS}1,\mathrm{DIS}2,\mathrm{DIS}3\},
\{\mathrm{F}1,\mathrm{F}2,\mathrm{F}3\},
\{\mathrm{DIS}^\ast,\mathrm{DF},\mathrm{F}^\ast\}
\bigr\}.
$
Our goal is to quantify the benefit of the consensus refinement itself, independently of the underlying base estimators. Accordingly, for each collection $\mathcal{A}$, we compare the refined output against the best individual estimator in $\mathcal{A}$ on the test split of \cite{cai2019dense}, which we denote by $A^\ast$.
Since we are not interested in comparisons between different $\mathcal{A}$, we measure performance by reporting the relative change in dataset-average \gls*{epe} with respect to this baseline. Let ${\mathrm{AEPE}}(A;\mathcal{D})$ be the average $\mathrm{EPE}$ attained by the algorithm $A$ on the dataset $\mathcal{D}$.
Then, we define the $\mathrm{rAEPE}(A,A^\ast)$ as
\begin{equation}
    \label{eq:average-epe-relative}
    \frac{{\mathrm{AEPE}}(A;\mathcal{D})
    -
    {\mathrm{AEPE}}(A^\ast;\mathcal{D})}
    {{\mathrm{AEPE}}(A^\ast;\mathcal{D})}
    \times 100~[\%].
\end{equation}
Negative values of $\mathrm{rAEPE}$ therefore indicate an improvement over the baseline $A^\ast$.

For the ablation studies, we use a separate $10~\%$ split of the train set from \cite{cai2019dense}, which we refer to as the validation set, and we tune the regularization parameters on an additional $10~\%$ split of the train set.
Often, \gls*{piv} involves outlier rejection and data interpolation schemes \cite{stamhuis2014pivlab}. To investigate the dependency of our framework on the quality of the outlier-rejection scheme and to assess its intrinsic data interpolation properties we define, for $\tau \geq 0$, $w_{i,\ell}^\tau$ as $w_{i,\ell}^\tau = w_{i,\ell}$ if $(\mathrm{\gls{epe}}_i)_{p(\ell)} \le \tau$, where $p(\ell)$ is the pixel entry corresponding to $\ell$, and $w_{i,\ell}^\tau = 0$ otherwise. We use the oracle outlier-rejection protocol as a controlled diagnostic experiment to decouple the evaluation of the consensus layer from any specific practical rejection heuristic. This setting reveals the performance attainable under progressively cleaner local estimates. As we show below, the results make clear that the proposed method remains effective even in the absence of any outlier-rejection mechanism. In \cref{subsec:experiments:outlier-rejection}, we also consider a simple learning-based rejection module and show that it can recover the oracle's performance. Because the development of a high-performance rejection scheme is not the main scope of this work, this experiment is intended mainly as a proof of feasibility beyond the oracle setting.

\subsection{Benefits of consensus}
\label{subsec:experiments:consensus}
In this section, we assess the benefits of the proposed method across different baseline algorithms sets $\mathcal{A}$. For reference, we also report results obtained by directly aggregating the different estimates using pixel-wise median and average.

\noindent\textbf{Experimental setup.}
For all $\mathcal{A}$, we report the $\mathrm{rAEPE}$ $[\%]$ \eqref{eq:average-epe-relative} 
on the test set for varying outlier-rejection threshold $\tau$. We consider $\phi(\cdot)$ in $f_i(\cdot)$ to be the Huber loss; see \cref{appendix:other-losses}. We tune the regularization parameters individually for each $\mathcal{A}$ as reported in \cref{appendix:regularizer-tuning}, and we set the weights $w_{i,\ell}$ as described in \cref{appendix:adjusted-pe}.

\begin{figure}
    \centering
    \includegraphics[width=8cm]{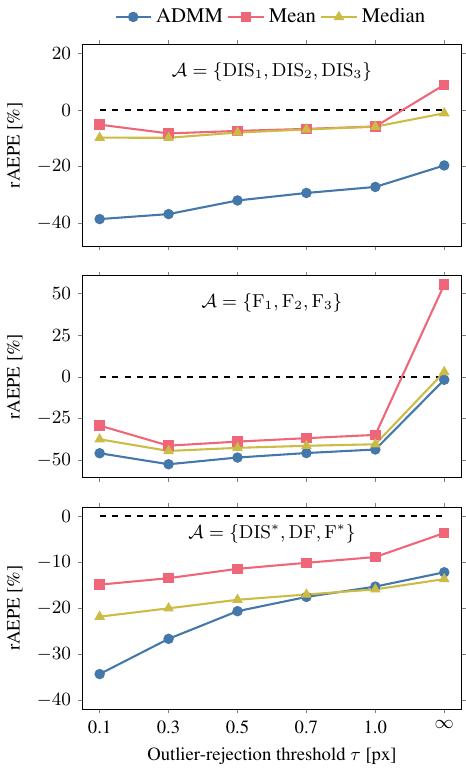}
    \vspace{-.25cm}
    \caption{Benefits of consensus across different sets of algorithms $\mathcal{A}$; see \cref{subsec:experiments:consensus}. We report the  $\mathrm{rAEPE}$ $[\%]$ \eqref{eq:average-epe-relative}
    (lower is better) over the outlier-rejection threshold~$\tau$ $[\text{px}]$.}
    \label{fig:consensus-benefits}
    \vspace{-.75cm}
\end{figure}

\noindent\textbf{Results.} 
We summarize the results in \cref{fig:consensus-benefits}. Notably, for all baseline algorithms $\mathcal{A}$, our consensus method yields a considerable performance improvement across all outlier-rejection thresholds. For instance, Farneb\"ack tops the $40~\%$ improvement with outlier rejection, and DIS the $20~\%$ across all $\tau$. Overall, these results suggest that a simple consensus layer on top of existing estimators may substantially improve their performance, and one can choose the estimators to balance accuracy and inference speed. Importantly, \gls*{admm}-based consensus also consistently outperforms simpler aggregation baselines such as pixel-wise mean and median.

\subsection{The effects of different \texorpdfstring{$f_i(\cdot)$}{data terms}}
\label{subsec:experiments:data-term}
An issue with \gls*{epe} as a metric, and consequently the consensus objective in \cref{subsec:distributed_data_term} is that regions of fast flow (corresponding to vectors with higher magnitude) are given more importance than slow regions. Thus, different data-terms $f_i(\cdot)$ may be more or less robust to different flows. We perform this ablation in this section.

\noindent\textbf{Experimental setup.} 
We fix $\mathcal{A} = \{\mathrm{DIS}1,\mathrm{DIS}2,\mathrm{DIS}3\}$, and for all the options of $f_i(\cdot)$ described in \cref{appendix:other-losses} ($\ell_1$ norm, squared $\ell_2$ norm, Huber loss) we report the $\mathrm{rAEPE}$ $[\%]$ \eqref{eq:average-epe-relative} on the validation set for varying outlier-rejection threshold $\tau$.
We tune the regularization parameters individually for each configuration as reported in \cref{appendix:regularizer-tuning}, and we set the weights $w_{i,\ell}$ as described in \cref{appendix:adjusted-pe}. For the $\ell_1$ loss, we use squared confidence weights. This reparameterization renders numerically comparable to the $\ell_2$ and Huber variants the influence
of the weights in the proximal update.

\noindent\textbf{Results.} 
\begin{figure}
    \centering
    \includegraphics[width=8cm]{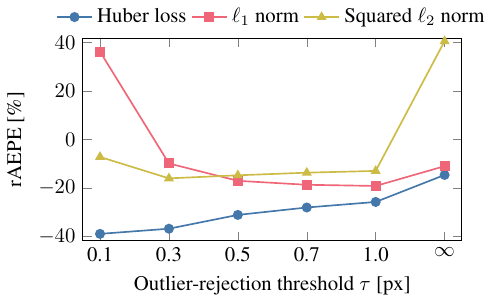}
    \vspace{-.25cm}
    \caption{Impact of data terms $f_i(\cdot)$; see \cref{subsec:experiments:data-term}. We report the  $\mathrm{rAEPE}$ $[\%]$ \eqref{eq:average-epe-relative} (lower is better) over the outlier-rejection threshold~$\tau$ $[\text{px}]$.}
    \label{fig:consensus-ablation-loss}
    \vspace{-.5cm}
\end{figure}
We report the results in \cref{fig:consensus-ablation-loss}. The Huber loss ranks as the most reliable: it consistently improves over the baseline across thresholds and degrades the least when $\tau$ is large, thanks to its built-in robustness to outliers. The $\ell_1$ norm is somewhat robust but can perform poorly when $\tau$ is very small (too few inliers), and the squared $\ell_2$ norm can work well only when outliers are already tightly filtered; without good rejection, it gets heavily corrupted by outliers and can make the consensus worse than the baseline.

\subsection{Weighting strategy}
\label{subsec:experiments:confidence}
In this section, we assess the impact of the different weighting strategies in the data term \eqref{eq:distributed_data_term}.

\noindent\textbf{Experimental setup.}
We fix $\mathcal{A} = \{\mathrm{DIS}1,\mathrm{DIS}2,\mathrm{DIS}3\}$ and compare different weighting strategies:
\begin{itemize}[leftmargin=*]
    \item \emph{Uniform averaging}: i.e., $w_{i,\ell} = 1$.
    \item \emph{$\mathrm{PE}$ weighting}: inverse photometric error as a confidence measure; cf. \cref{appendix:pe}.
    \item \emph{(Gradient-)Adjusted $\mathrm{PE}$ weighting}: inverse photometric error adjusted based on the image texture; cf. \cref{appendix:adjusted-pe}.
    \item \emph{(Gradient-)Adjusted uniform averaging}: uniform confidence over all algorithms adjusted based on the image texture; cf. \cref{appendix:adjusted-uniform}.
\end{itemize}
In all cases, we consider the Huber loss (cf. \cref{appendix:other-losses}) and we tune the regularization parameters individually for each configuration as reported in \cref{appendix:regularizer-tuning}. We report the $\mathrm{rAEPE}$ $[\%]$ \eqref{eq:average-epe-relative} on the validation set for varying outlier-rejection threshold $\tau$.

\noindent\textbf{Results.}
\begin{figure}
    \centering
    \includegraphics[width=8.75cm]{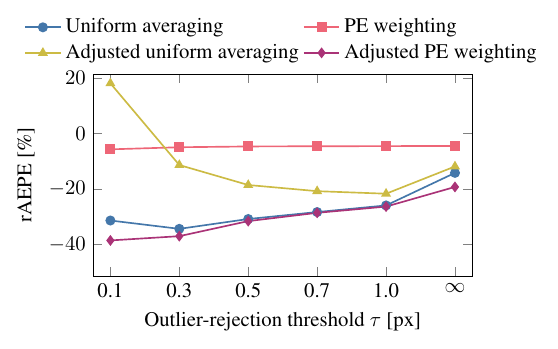}
    \vspace{-.75cm}
    \caption{Impact of weighting schemes; see \cref{subsec:experiments:confidence}. We report the  $\mathrm{rAEPE}$ $[\%]$ \eqref{eq:average-epe-relative} (lower is better) over the outlier-rejection threshold~$\tau$ $[\text{px}]$.}
    \label{fig:consensus-ablation-weighting}
    \vspace{-.25cm}
\end{figure}
We summarize the results in \cref{fig:consensus-ablation-weighting}. The results highlight the benefits of the gradient-adjusted $\mathrm{PE}$ weighting scheme over the naive $\mathrm{PE}$ weighting, as well as the perhaps surprising efficacy of simple uniform averaging.
\subsection{Effects of regularization}
\label{subsec:experiments:regularization}

In this section, we assess the effects ofregularization \eqref{eq:reg_sum}.

\noindent\textbf{Experimental setup.}
We fix $\mathcal{A} = \{\mathrm{DIS}1,\mathrm{DIS}2,\mathrm{DIS}3\}$ and consider $\phi(\cdot)$ in $f_i(\cdot)$ to be the Huber loss; see \cref{appendix:other-losses}. To understand the effects of regularization independently of the other elements of the pipeline, we also consider a fictitious weighting strategy in which $w_{i,\ell} = 1$ if and only if algorithm $A_i$ provided the best estimate at the corresponding location, i.e., 
\begin{equation}
w_{i,\ell} =
\begin{cases}
    1&\text{if }i = \underset{j \in \{1,2,3\}}{\argmin} \|\groundtruthflow(x_\ell, y_\ell) - \startflow_j(x_\ell, y_\ell)\|_2^2\\
    0&\text{otherwise},
\end{cases}
\label{eq:ideal-weighting}
\end{equation} 
where $\groundtruthflow$ is the ground-truth, $(x_\ell, y_\ell)$ is the pixel corresponding to the $\ell$-th entry, and we consider an arbitrary tie-breaking strategy so that the $\argmin$ is a singleton. 
Overall, we compare four methods:
\begin{itemize}[leftmargin=*]
    \item $w_{i,\ell}$ as in \cref{appendix:adjusted-pe}, without regularization ($\lambda_\bullet = 0$).
    \item $w_{i,\ell}$ as in \cref{appendix:adjusted-pe}, with regularization.
    \item $w_{i,\ell}$ as in \eqref{eq:ideal-weighting}, without regularization ($\lambda_\bullet = 0$).
    \item $w_{i,\ell}$ as in \eqref{eq:ideal-weighting}, with regularization.
\end{itemize}
When using regularization, we select $\lambda_\bullet$ as in \cref{appendix:regularizer-tuning}.

\begin{figure}
    \centering
    \includegraphics[height=5cm]{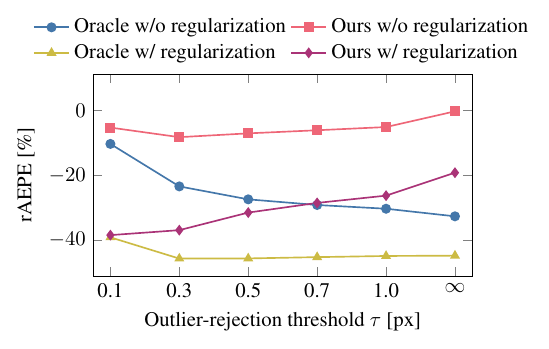}
    \vspace{-.25cm}
    \caption{Impact of the regularization; see \cref{subsec:experiments:regularization}. We report the  $\mathrm{rAEPE}$ $[\%]$ \eqref{eq:average-epe-relative} (lower is better) over the outlier-rejection threshold~$\tau$ $[\text{px}]$.}
    \label{fig:consensus-ablation-regularization}
    \vspace{-.25cm}
\end{figure}

\noindent\textbf{Results.} 
We summarize the results in \cref{fig:consensus-ablation-regularization}. Overall, all the results consistently show the benefit of regularization, as well as the benefit of better weighting schemes: The ideal weighting suggests the presence of additional performance gains, motivating future work to investigate different, and possibly learning-based, schemes.

\subsection{Learning-based outlier rejection}
\label{subsec:experiments:outlier-rejection}

Next, we move beyond the oracle protocol and show that the proposed consensus framework can be combined with a practical learning-based outlier-rejection module.

\noindent\textbf{Experimental setup.}
We consider the DIS and Farneb\"ack estimator families separately. For each family, we train a U-Net \cite{unet} to predict, at each pixel, if the local estimate has endpoint error larger than $1$ pixel. At inference time, these predictions are used to reject unreliable estimates in the consensus procedure. We compare three consensus configurations: without outlier rejection, with oracle rejection at threshold $\tau=1$, and with the learned rejection module.

\noindent\textbf{Results.}
We report the results in \cref{tab:learning-based-outlier-rejection}. The learning-based rejection module nearly recovers the gain obtained with oracle rejection at $\tau=1$, showing that a practical scheme can capture most of the benefit of oracle filtering.

\begin{table}
    \centering
    \begin{tabular}{lcc}
        \toprule
         & $\{\mathrm{DIS}1,\mathrm{DIS}2,\mathrm{DIS}3\}$ & $\{\mathrm{F}1,\mathrm{F}2,\mathrm{F}3\}$ \\
        \midrule
        W/o rejection            & 19.63 & 1.8253 \\
        Oracle ($\tau=1$)      & 27.24 & 43.623 \\\hline
        \rowcolor{blue!5} U-net      & 26.89 & 43.568 \\
        \bottomrule
    \end{tabular}
    \caption{Baseline and consensus performance with no rejection, oracle rejection ($\tau=1$), and learning-based rejection; see \cref{subsec:experiments:outlier-rejection}. We report $\mathrm{rAEPE}$ $[\%]$ \eqref{eq:average-epe-relative}.}
    \label{tab:learning-based-outlier-rejection}
    \vspace{-.5cm}
\end{table}
\section{Active fluid control}
\label{sec:afc}

We conclude by assessing if the proposed method can be deployed in a real active-fluids-control loop.

\noindent\textbf{Experimental setup.}
We use exactly the same active-fluids-control setup, tasks, observations, rewards, and training protocol as in~\cite{terpin2025ff}. In particular, we consider both drag minimization and drag maximization on the spinning-cylinder system, and we repeat the training three times for each task. The only change is the flow-quantification module used to construct the flow-based observation provided to the agent: we use the proposed consensus \gls*{admm}. Within this framework, the local sub-estimators are given by the same DIS estimator as in~\cite{terpin2025ff}, together with variants using smaller and larger patch sizes and different regularization settings.

\begin{figure}
    \centering
\includegraphics[width=8cm]{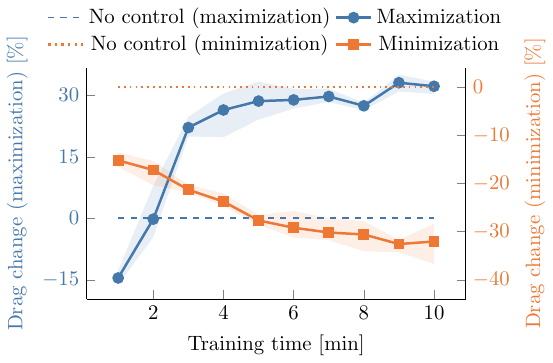}
\vspace{-.25cm}
    \caption{Training curves for the active-fluids-control tasks of drag maximization and drag minimization. The blue curve reports performance on the maximization task (left axis), and the orange curve reports performance on the minimization task (right axis). Solid and dashed lines show the mean across runs, and the shaded bands indicate the run-to-run range. The horizontal zero lines denote the no-control baseline.}
    \label{fig:afc_learning_curves}
    \vspace{-.5cm}
\end{figure}

\noindent\textbf{Results.}
We report the corresponding training curves in \cref{fig:afc_learning_curves} for both drag minimization and drag maximization. In both cases, the agent can be trained successfully using flow observations generated by the proposed method, achieving drag reductions of $36~\%$ and drag increases of up to $32~\%$ with only two minutes of real-world interaction. These results show that the consensus-based refinement is compatible with closed-loop operation in the real system and can be deployed successfully to learn effective active-fluids-control policies.

\section{ACKNOWLEDGMENTS}
The authors are grateful to Florian D\"orfler for providing feedback on a preliminary version of this work, and Raff D'Andrea for the infrastructure, funding and guidance. This work was supported by a ETH RobotX research grant funded through the ETH Zurich Foundation.

\bibliographystyle{IEEEtran}
\bibliography{bib}

\appendix
\subsection{Choices of the data term\texorpdfstring{ $f_i(\cdot)$}{}}
\label[appendix]{appendix:other-losses}

In this subsection, we derive the reformulations of \eqref{eq:local-update} when \eqref{eq:distributed_data_term} is the \texorpdfstring{$\ell_1$}{l1}, \texorpdfstring{$\ell_2$}{l2} norm or the Huber loss \cite{huber1992robust}.

\noindent\textbf{\texorpdfstring{$\ell_1$}{l1} norm.}
The proximal operator of the $\ell_1$ norm is a soft-thresholding and, thus, \eqref{eq:local-via-proximal} becomes:
\begin{equation*}
 (\startflow_i)_\ell
+
\mathrm{sign}(r)\max\left(|r| - {w_{i,\ell}}\rho^{-1}, 0\right).
\label{eq:l1:closedform}
\end{equation*}

\noindent\textbf{Squared \texorpdfstring{$\ell_2$}{l2} norm.}
For the squared $\ell_2$ norm, \eqref{eq:local-via-proximal} becomes:
\begin{equation*}
{2w_{i, \ell} \startflow_i + \rho(\optVar[k] - \dual_i[k])}(2w_{i, \ell} + \rho)^{-1}.
\label{eq:l2:closedform}
\end{equation*}

\noindent\textbf{Huber loss.}
For $\delta \geq 0$, the Huber loss $\phi_\delta(t)$, defined as $\phi_\delta(t) = \dfrac{1}{2}t^2$ if $|t|\le\delta$ and $\phi_\delta(t) = \delta\bigl(|t|-\tfrac{1}{2}\delta\bigr)$ otherwise, admits the proximal operator \cite{parikh2014proximal} for $\lambda \geq 0$:
\[
\operatorname{prox}_{\lambda \phi_\delta}(v)
=
\begin{cases}
{v}(1+\lambda)^{-1},
& \text{if }|v|\le (1+\lambda)\delta,\\
v - \lambda \delta,
& \text{if }v > (1+\lambda)\delta,\\
v + \lambda \delta,
& \text{if }v < -(1+\lambda)\delta.
\end{cases}
\]
In our experiments, we use $\delta = 1$. The update in \eqref{eq:local-via-proximal} is then computed with $v = w_{i, \ell}r$ and $\lambda = 1/\rho$.

\subsection{Confidence weighting}
\label[appendix]{appendix:confidence}
Here, we discuss three notions of per-pixel confidence.

\subsubsection{Photometric error}
\label[appendix]{appendix:pe}
In optical flow literature \cite{baker2004lucas,kroeger2016fast}, a common proxy for local estimation quality is the photometric error. The photometric error measures the average residual of the brightness constancy equation, $\left(I_0(x,y) - I_1(x+\startflow_{i,x}(x, y),y+\startflow_{i,y}(x, y)) \right)^2$, in a patch $P_{(\bar{x}, \bar{y})}$ centered at pixel $(\bar{x}, \bar{y})$ corresponding to the entry $\ell$ (note that there are two values of $\ell$ corresponding to the same $(x, y)$ in our notation) and $(\startflow_{i,x}(x, y),\startflow_{i,y}(x, y))$ is the estimate of algorithm $i$ at the pixel $(x, y)$.
We then define
\begin{equation}
\label{eq:photometric_error}
    w_{i,\ell} = \sqrt{\mathrm{PE}_{i}(\bar{x}, \bar{y})}^{-1}.
\end{equation}
This metric captures how well an estimate satisfies brightness constancy, but it does not reflect the distinctiveness of the region: in areas with low image gradient magnitude, many different displacements can produce similarly low residuals. 

\subsubsection{Gradient-adjusted photometric error}
\label[appendix]{appendix:adjusted-pe}
Photometric error alone does not capture local identifiability: in low-texture regions, many displacements can yield similarly small residuals. Motivated by this observation, we define
\begin{equation}
    w_{i,\ell} = {\|\nabla I_0(x, y)\|_2}\sqrt{\mathrm{PE}_{i}(x, y)}^{-1},
    \label{eq:pixel_precision}
\end{equation}
where $(x, y)$ is the pixel corresponding to the entry $\ell$.
Intuitively, the expression in \eqref{eq:pixel_precision} increases confidence where the estimator achieves a low photometric error and the starting image has a high gradient magnitude. On the other hand, the confidence decreases where the fit to the brightness constancy equation is poor or the original image has low features. 
This practical weighting rule is heuristic. Its scaling is motivated by an idealized variance calculation whose assumptions need not hold for the \gls*{piv} algorithms used here. Regularization, coarse-to-fine processing, estimator bias in shear layers, anisotropic covariance, and higher-order effects may therefore limit its validity. Accordingly, the calculation is neither a calibrated uncertainty model nor a performance guarantee; addressing these effects remains future work.
\newlist{assumptionenum}{enumerate}{1}
\setlist[assumptionenum,1]{label*=A\theassumption.\arabic*}
\begin{assumption}
\label{assumption:blue}
Let $\startflow_i(x,y), \groundtruthflow(x, y) \in \mathbb{R}^2$ be the estimate of the $i$-th algorithm and the ground truth, respectively, evaluated at pixel $(x,y)$. We assume that:
\begin{assumptionenum}
    \item The brightness constancy assumption holds, and is thus satisfied by the ground-truth flow $\groundtruthflow$. \label{ass:bright-const}
    \item The expected value $\mathbb{E}[\startflow_i(x,y)]$ of the estimates is unbiased for all $i$, $(x,y)$; i.e., $\mathbb{E}[\startflow_i(x,y)]=\groundtruthflow(x, y)$. \label{ass:unbiased}
    \item The covariance matrix of the estimates satisfies $\Var[\startflow_i(x,y)] = \sigma^2_i(x,y)\mathbb{I}_2$ for all $i$, $(x,y)$\footnote{Here, we denote the two-dimensional identity matrix via $\mathbb{I}_2$.}.\label{ass:uncorrel_xy}
\end{assumptionenum}
\end{assumption}

We can derive an upper bound on $\sigma^2_i(x,y)$:

\begin{proposition}
\label[proposition]{prop:upper-bound-pe}
Let $\startflow_i(x,y), \groundtruthflow(x,y) \in \mathbb{R}^2$ be the estimate of the $i$-th algorithm and the ground truth, respectively, evaluated at pixel $(x,y)$. If \cref{assumption:blue} holds and $\nabla I_0(x,y)\neq 0$, then the variance of the pixel estimates $\sigma_i^2(x,y)$ (cf. \cref{ass:uncorrel_xy}) is upper bounded as
\begin{equation}
\label{eq:upper-bound-pe}
\sigma_i^2(x,y) \leq {\mathbb{E}\left[e^2 + 2C\|g\|\|\Delta\|^3\right]}\|g\|^{-2},
\end{equation}
where $g = \nabla I_0(x,y)$, $\Delta = \Delta \startflow_i(x,y)$, $C\ge 0$, and 
$
e = I_0(x,y) - I_1(x+\startflow_{i,x}(x,y), y+\startflow_{i,y}(x,y)).
$
\end{proposition}

\begin{proof}
Each algorithm $A_i$ produces an approximate solution $\startflow_i(x,y)$ with residual $e$.
By \cref{ass:bright-const}, using the inverse compositional formulation \cite{bakerimagealignment} yields
$
I_0(x-\startflow_{i,x}(x,y), y-\startflow_{i,y}(x,y)) - I_1(x,y) = e
$
and
$I_0(x-\groundtruthflow_x(x,y), y-\groundtruthflow_y(x,y)) - I_1(x,y) = 0.$
Subtracting these identities gives
$I_0(x-\groundtruthflow_x(x,y), y-\groundtruthflow_y(x,y))
-
I_0(x-\startflow_{i,x}(x,y), y-\startflow_{i,y}(x,y))
= -e.$
Linearizing $I_0$ around $(x,y)$, we obtain $g^\top \Delta = -e + r$, with $|r|\le C\|\Delta\|^2$
for some $C\ge 0$.
Hence,
$
e^2 
= (g^\top\Delta)^2 - 2rg^\top\Delta + r^2
\ge (g^\top\Delta)^2 - 2|r||g^\top\Delta|.
$
Using $|r|\le C\|\Delta\|^2$ and $|g^\top\Delta|\le \|g\|\|\Delta\|$, we get
$
e^2 \ge (g^\top\Delta)^2 - 2C\|g\|\|\Delta\|^3.
$
Taking expectations yields
$
\mathbb E[(g^\top\Delta)^2]
\le
\mathbb E[e^2] + 2C\|g\|\mathbb E[\|\Delta\|^3].
$
For constant $\groundtruthflow(x,y)$, \cref{ass:unbiased} gives $\mathbb E[\Delta]=0$. Therefore,
$\mathbb E[(g^\top\Delta)^2]
=
g^\top \Var[\Delta] g.$
Moreover, for constant $\groundtruthflow(x,y)$, $I_0(x,y)$, and $I_1(x,y)$, we have
$\Var[\Delta] = \Var[\startflow_i(x,y)].$
Thus, by \cref{ass:uncorrel_xy}, $g^\top \Var[\Delta] g
=
\sigma_i^2(x,y)\|g\|^2$.
Therefore,
$
\sigma_i^2(x,y)\|g\|^2
\le
\mathbb E[e^2] + 2C\|g\|\,\mathbb E[\|\Delta\|^3].
$
We conclude dividing by $\|g\|^2$.
\end{proof}

The expected value $\mathbb{E}[e_i^2]$ can be computed using the one-sample estimate:
$
\mathbb{E}[e_i^2] \approx e_i^2 = \mathrm{PE}_i(x,y).
$
Motivated by \cref{prop:upper-bound-pe}, and under the heuristic assumption that the higher-order term is small, we approximate
\begin{equation}
    w_{i,\ell} = \dfrac{1}{\sigma_{i}}
    \approx \dfrac{\|\nabla I_0(x,y)\|_2}{e_i} \approx \dfrac{\|\nabla I_0(x,y)\|_2}{\sqrt{\mathrm{PE}_i(x,y)}}.
    \label{eq:pixel_confidence}
\end{equation}

\subsubsection{Gradient-adjusted uniform weighting}
\label[appendix]{appendix:adjusted-uniform}
When $\lambda_\bullet = 0$, \eqref{eq:photometric_error} and \eqref{eq:pixel_confidence} yield the same consensus variables given the same initial estimates, as there is no coupling across pixels. For this reason, inspired by this similarity, we also consider a third additional weighting scheme that adjusts the uniform scheme with gradient information $w_{i,\ell} = \|\nabla I_0(x,y)\|_2$.

\subsection{Regularization hyperparameters}
\label[appendix]{appendix:regularizer-tuning}
Without any aim of optimal hyperparameter tuning, we tune $\lambda_\bullet$ by starting with $\lambda_\bullet = 0$ and increasing, in order and as long as the performances on the validation set (we use $10~\%$ of the train set in \cite{cai2019dense}) improve, $\lambda_{\mathrm{div}}, \lambda_{\mathrm{curv}}, \lambda_{\mathrm{s}}$:
\begin{center}
\begin{tabular}{lccc}
\toprule
\multicolumn{4}{c}{$\mathcal{A} = \{\mathrm{DIS}1, \mathrm{DIS}2, \mathrm{DIS}3\}$, Huber 
}
\\
$w_{i,\ell}$
& $\lambda_{\mathrm{s}}$
&
$\lambda_{\mathrm{curv}}$
&
$\lambda_{\mathrm{div}}$\\\hline
$w_{i,\ell} = 1$ &
$0$ & {$5$} & {$300$} \\
$w_{i,\ell}$ as in \cref{appendix:pe} & {$0$} & {$0.2$} & {$1$} \\
$w_{i,\ell}$ as in \cref{appendix:adjusted-pe} & {$3$} & {$30$} & {$1300$} \\
$w_{i,\ell}$ as in \cref{appendix:adjusted-uniform} & {$200$} & {$550$} & {$750$} \\
$w_{i,\ell}$ as in \eqref{eq:ideal-weighting} & {$0$} & {$1$} & {$10$} \\
\hline\hline
\multicolumn{4}{c}{$\mathcal{A} = \{\mathrm{DIS}1, \mathrm{DIS}2, \mathrm{DIS}3\}$, $w_{i,\ell}$ as in \cref{appendix:adjusted-pe}}\\
$\phi(\cdot)$
& $\lambda_{\mathrm{s}}$
&
$\lambda_{\mathrm{curv}}$
&
$\lambda_{\mathrm{div}}$\\\hline
    $\ell_1$
    & {$150$} & {$250$} & {$1100$} \\
    $\ell_2$
    & {$5$} & {$500$} & {$1000$} \\
    Huber
    & {$3$} & {$30$} & {$1300$} \\\hline\hline
\multicolumn{4}{c}{Huber%
, $w_{i,\ell}$ as in \cref{appendix:adjusted-pe}}\\
$\mathcal{A} $ & $\lambda_{\mathrm{s}}$
&
$\lambda_{\mathrm{curv}}$
&
$\lambda_{\mathrm{div}}$\\\hline
$\mathcal{A} = \{\mathrm{F}1, \mathrm{F}2, \mathrm{F}3\}$& {$15$} & {$20$} & {$700$} \\
$\mathcal{A} = \{\mathrm{DIS}^\ast, \mathrm{DF}, \mathrm{F}^\ast\}$ & {$0$} & {$25$} & {$1200$} \\\bottomrule
\end{tabular}
\end{center}

\end{document}